\documentclass[reprint, amsmath,amssymb, aip, apl]{revtex4-2}
\usepackage{booktabs} 
\usepackage{eurosym}
\usepackage{graphicx}
\usepackage{dcolumn}
\usepackage{bm}
\usepackage{graphicx}
\usepackage[colorlinks=true, allcolors=blue]{hyperref}
\usepackage[capitalize,nameinlink]{cleveref}
\usepackage{subcaption}
\usepackage{caption}
\usepackage{ragged2e}
\usepackage{overpic}

\begin{document}

\preprint{APS/123-QED}

\title{Quantum Transport and Apparent Work Function Distributions of Atomic Contacts via a 3D-Printed High-Vacuum Platform}
\author{G. Pellicer}
\affiliation{Departamento de Física and Instituto Universitario de Materiales de Alicante (IUMA), Universidad de Alicante, E-03690 Alicante, Spain.}
\author{C. Sabater}
 \email{carlos.sabater@ua.es}
\affiliation{Departamento de Física and Instituto Universitario de Materiales de Alicante (IUMA), Universidad de Alicante, E-03690 Alicante, Spain.}

\date{\today}

\begin{abstract}
We present a low-cost, 3D-printed high-vacuum platform integrating a mechanically controllable break-junction system and a custom logarithmic amplifier for room-temperature quantum transport measurements. Using copper as a highly reactive test case, we successfully resolve the $1G_0$ conductance quantum under both high vacuum and anhydrous glycerol, demonstrating the effectiveness of these environments against rapid atmospheric oxidation. In parallel, utilizing gold as a robust benchmark, we systematically extract the apparent work function ($\phi$) from thousands of tunneling traces across ambient air, vacuum, and glycerol. Our analysis demonstrates that the statistical distribution of $\phi$ rigorously follows a non-central chi-square distribution. The obtained gold work functions match existing literature across all environments. Although lower than bulk values, they perfectly align with theoretical models accounting for atomic-scale roughness, apex geometry, and environmental adsorbates. Ultimately, this methodology establishes an accessible and reproducible framework for systematic nanoscale research on reactive materials.
\end{abstract}

\maketitle


\section{INTRODUCTION}

As classical microelectronics reaches its physical limits~\cite{moore1965cramming, dennard1974design, huang2022gtc, gelsinger2023innovation, intel2024angstrom}, molecular electronics emerges as an  alternative to transcend conventional lithographic constraints~\cite{aviram1974molecular, Cuevasbook, natelson2015nanostructures, Oliver23}. This single-molecule approach serves as a fundamental benchmark for studying in-depth quantum transport and developing decoherence-resistant qubits~\cite{Yao2025, gaita19, Lapobogani08}. However, its ultimate success relies on achieving stable charge transport at ambient temperatures, overcoming the restrictive and expensive requirements of cryogenic environments~\cite{Guedon12}.

To create atomic or single-molecule junctions, the most prevalent approaches are Scanning Tunneling Microscopy Break Junctions (STM-BJ)\cite{Pascual1993,XuTao2003} and Mechanically Controllable Break Junctions (MCBJ)\cite{muller1992, Krans93}. These techniques operate by repeatedly forming and breaking metallic contacts.  Quantum transport at the atomic scale is described by the Landauer formalism~\cite{Landauer57} $G = G_0 \sum_i T_i$. Here, $T_i$ represents the transmission probability of the $i$-th conduction channel, and $G_0 = \frac{e^2}{\pi\hbar}$ denotes the conductance quantum (where $\hbar$ is the reduced Planck constant) \cite{Landauer57, buttiker1986four, vanwees1988quantized, wharam1988one}. For noble metals like gold (Au), silver (Ag), and copper (Cu), the outer \textit{s}-valence orbital results in a single open channel ($\sum_i T_i \approx 1$), yielding $G \approx G_0 \approx 12\,906~\Omega^{-1}$~~\cite{Agrait2003}. Upon further elongation and rupture of this atomic contact or molecular bridge, the system transitions into the tunneling regime. In this state, the conductance decays exponentially with the relative electrode displacement $z$ according to:
\begin{equation}
G(z) = G_i \exp\left(-\frac{2\sqrt{2m_e\phi}}{\hbar}z\right)
\label{eq:tunneling}
\end{equation}
where $m_e$ is the electron mass and $G_i$ is the initial tunneling conductance. This exponential decay provides a direct method to extract the apparent work function ($\phi$).

While these conductance quantum and tunneling features are easily observable in Au even under ambient conditions~\cite{Borja20, Pellicer}, gold's high work function ($\phi_\text{Au} = 5.1{-}5.4$~eV)~\cite{Michaelson77} often misaligns with molecular HOMO levels ($4.5$ to $5.5$~eV) typical of organic electronics. Copper ($\phi_\text{Cu} = 4.5{-}5.0$~eV)~\cite{Michaelson77}, offering optimal HOMO-$E_F$ alignment and enhanced $d$-orbital coupling to molecular $\pi^*$ states~\cite{Venkataraman2006}, represents the ideal electrode material for molecular electronics. However, copper and alternative metals such as platinum~\cite{Yelin2016} or aluminum exhibit high reactivity with oxygen, leading to rapid oxide formation that severely obscures quantum transport~\cite{CzyszczonBurton2025Single}. Furthermore, the apparent work function ($\phi_\text{Au}$) extracted from the tunneling regime is fundamentally lower\cite{Gimzewski} than the macroscopic $5.3$~eV. This intrinsic reduction originates from the extreme atomic-scale roughness at the contact apex, where the Smoluchowski smoothing of the electron cloud creates a local dipole that lowers the potential barrier~\cite{Smoluchowski}, a geometric modification of the apparent barrier characteristic of tunneling junctions~\cite{LangD88}. In ambient conditions, this already reduced barrier is drastically exacerbated, dropping down to $\approx 1$~eV due to molecules adsorbing onto the metal surface~\cite{VanWee2005}.

This environmental degradation represents a critical bottleneck for MCBJ techniques: transport measurements in the tunneling regime are compromised by electrical and mechanical noise and atmospheric contamination. Traditionally, preventing oxidation or contamination requires expensive and complex vacuum~\cite{Weber69,lowWF2025} or cryogenic equipment~\cite{Klaker24,kim10}.

In this work, we address both the physical and economic barriers of characterizing reactive metals at room temperature. We introduce a fully integrated 3D-printed polylactic acid (PLA) platform that encapsulates a custom MCBJ setup. The assembled chamber withstands atmospheric pressure to reliably reach high-vacuum conditions. Furthermore, to seamlessly capture the transition from the atomic contact to the tunneling regime, we implemented a custom-designed logarithmic amplifier. This circuitry spans a continuous dynamic range down to $10^{-4} G_0$ without interruptions or data loss, overcoming the challenge of accurate statistical work function extraction. We validate this platform through two key demonstrations. First, we preserve and resolve the $1G_0$ conductance quantum of highly reactive Cu atomic contacts, an achievement that is typically impossible in ambient air. Second, utilizing gold as a resilient benchmark, we conduct a comprehensive statistical extraction of the apparent work function across diverse environments (ambient air, high vacuum, and immersion in anhydrous glycerol). This statistical approach quantifies how atmospheric adsorbates drastically alter surface energetics. Ultimately, this methodology establishes a robust, highly affordable, and open-source framework for systematic nanoelectronics research on reactive materials.

\section{Experimental Setup: 3D-Printed Hardware and Custom Electronics}

The experimental methodology presented in this work is driven by a primary technological challenge: demonstrating that a simple, low-cost vacuum chamber constructed via filament-based 3D printing can reliably achieve and sustain high-vacuum conditions. Consequently, this section first details the design and physical validation of this custom PLA chamber\cite{Spicer}. Following this proof of concept, we describe the fundamentals of the MCBJ technique, its specific 3D-printed integration within the encapsulated environment, and finally, the implementation of a custom logarithmic amplifier tailored for high-resolution electronic transport measurements.

\subsection{3D-Printed Vacuum Chamber} \label{sec:vacuum_chamber}

In this work, we demonstrate that filament-based 3D printing is a viable and highly effective method for fabricating functional vacuum chambers. Utilizing standard PLA filament significantly reduces both production costs and manufacturing time compared to traditional machining processes. Furthermore, because PLA is derived from renewable resources such as corn starch or sugarcane, this approach offers a sustainable, eco-friendly alternative for developing custom laboratory instrumentation.

The high-vacuum chamber and the internal MCBJ structural components were fabricated using Fused Deposition Modeling (FDM) on a commercial desktop 3D printer (Creality Ender-3 S3)\cite{crealityender}. To ensure structural integrity and strict hermeticity against the 1 atm pressure differential, all parts were printed using PLA filament with a 100\% infill density and a layer height of 0.12 mm. The extrusion temperature was maintained at 210~$^\circ$C to promote optimal layer adhesion and minimize microscopic voids, while the heated bed was set to 60~$^\circ$C to prevent thermal warping during the printing process.

\begin{figure}[h!]
            
        \centering
        \includegraphics[width=0.99\linewidth]{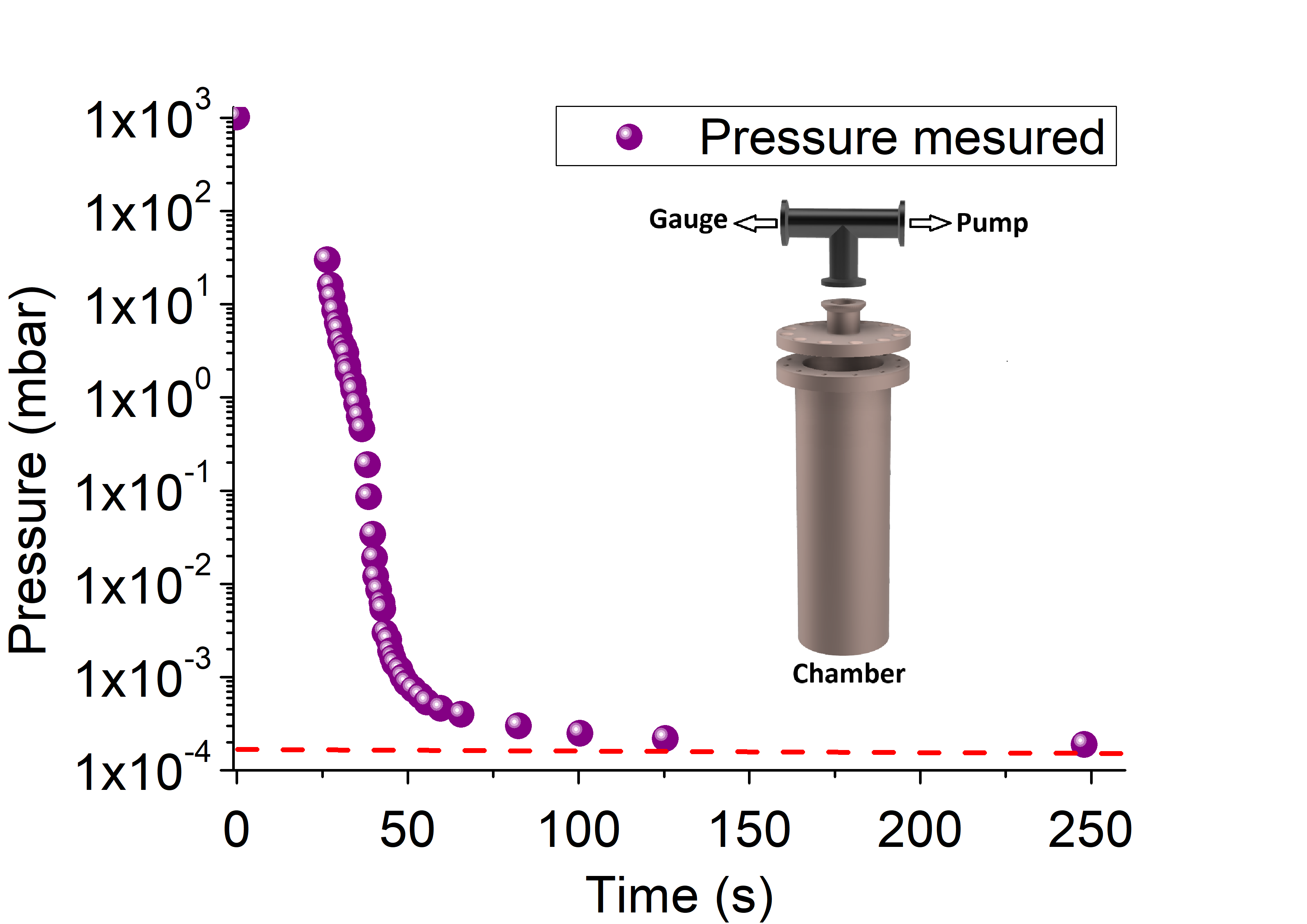}
  \captionsetup{justification=justified, singlelinecheck=false}
    \caption{\justifying Vacuum pump-down curve of the 3D-printed system over time. The dashed red line indicates the asymptotic high-vacuum limit ($1.4 \times 10^{-4}$~mbar) achieved by the setup. Inset: CAD model of the PLA chamber showcasing the integrated KF flange for standard vacuum line connections.}
    \label{fig:chamber}
\end{figure}

Since the pressure differential between the high-vacuum environment and the laboratory atmosphere is strictly limited to one atmosphere, standard PLA possesses sufficient mechanical integrity for this application. Our design features a cylindrical chamber with a side aperture surrounded by a holed ring, which mounts directly onto the base containing the experimental setup. To ensure a reliable, hermetic connection, the chamber nozzle incorporates an integrated, 3D-printed KF flange, enabling a seamless interface with standard commercial vacuum lines (see inset Fig.~\ref{fig:chamber}).

For experimental validation, the 3D-printed chamber was connected to a high-vacuum system consisting of a turbomolecular pump backed by a dry diaphragm pump~\cite{hicube80}. The system successfully reached and maintained a high-vacuum regime of $1.4 \times 10^{-4}$~mbar within 2 hours (dashed red line). However, as illustrated by the pump-down curve in Fig.~\ref{fig:chamber}, even with conventional, low-power vacuum equipment, the pressure rapidly dropped to its stable vacuum plateau in less than 2 minutes. The absence of significant leaks demonstrates that this 3D-printed architecture is a robust, functional, and accessible alternative to conventional metal-based high-vacuum systems.

\subsection{MCBJ Fundamentals and 3D Integration}\label{sec:mcbj_description}

At the core of the experimental methodology is the MCBJ mechanism, designed to generate and stabilize atomic-sized contacts through the controlled rupture of a notched metallic wire. In this configuration, the wire is mounted onto a flexible, 3D-printed PLA substrate~\cite{JPCuenca26,Pellicer}. For these experiments, Au and Cu wires (Goodfellow, $99.998\%$ purity, $0.25$~mm diameter) were utilized. A central notch was precisely fabricated on the wires using a guillotine-cut method to localize the mechanical stress. 

Electrical characterization is performed by applying a constant bias voltage ($V_{\mathrm{bias}}$) across the junction. Mechanical control is managed via a computer-generated sawtooth waveform, which is fed into a high-voltage amplifier to drive a piezoelectric actuator. This actuator bends the flexible substrate, enabling the precise opening and closing of the junction with sub-angstrom resolution. To ensure accurate displacement measurements, the relative distance between the electrodes was calibrated following the protocol detailed in Ref.~\cite{JPCuenca26}. 


This continuous mechanical cycling enables the real-time recording of the transport signal. The measured current is processed by a custom logarithmic current-to-voltage ($I$--$V$) amplifier~\cite{escorza2026,ILOGAZhang}, digitized by a data acquisition (DAQ) system~\cite{ni6363}, and mathematically converted into $G_0$ units. The resulting curves are fundamentally classified into two categories: rupture traces (acquired as the junction is elongated and broken) and formation traces (acquired as the atomic contact is re-established)~\cite{Agrait2003}. Furthermore, following established methodologies that utilize liquid environments to protect atomic-scale interfaces~\cite{XuTao2003, Zhang2025}, a drop of glycerol ($99\%$ purity, Panreac AppliChem ITW Reagents) was deposited directly onto the notched region via drop-casting to act as an effective shield against environmental contamination.

The core of our custom setup features a dedicated MCBJ system encapsulated within a 3D-printed PLA vacuum chamber. This integrated approach enables rapid prototyping and cost-effective customization without compromising the structural integrity needed to withstand pressure differentials. To achieve a hermetic seal for stable high-vacuum conditions, the chamber is secured to the main experimental body using an indium wire seal. Additionally, modular quick-connect fittings are integrated into the design and permanently sealed with Stycast 2850FT \cite{Stycast2850FT}. These fittings grant the system significant versatility, allowing it to be efficiently evacuated or backfilled with controlled inert atmospheres, such as Helium (He) or Argon (Ar).

\begin{figure}[htp]
    \centering
    \captionsetup[subfigure]{position=top, justification=raggedright, singlelinecheck=false, labelfont=normalsize}

    \begin{subfigure}{\linewidth}
        \caption{} \label{fig:setupMCbJa}
        \centering
         \includegraphics[width=0.9\textwidth]{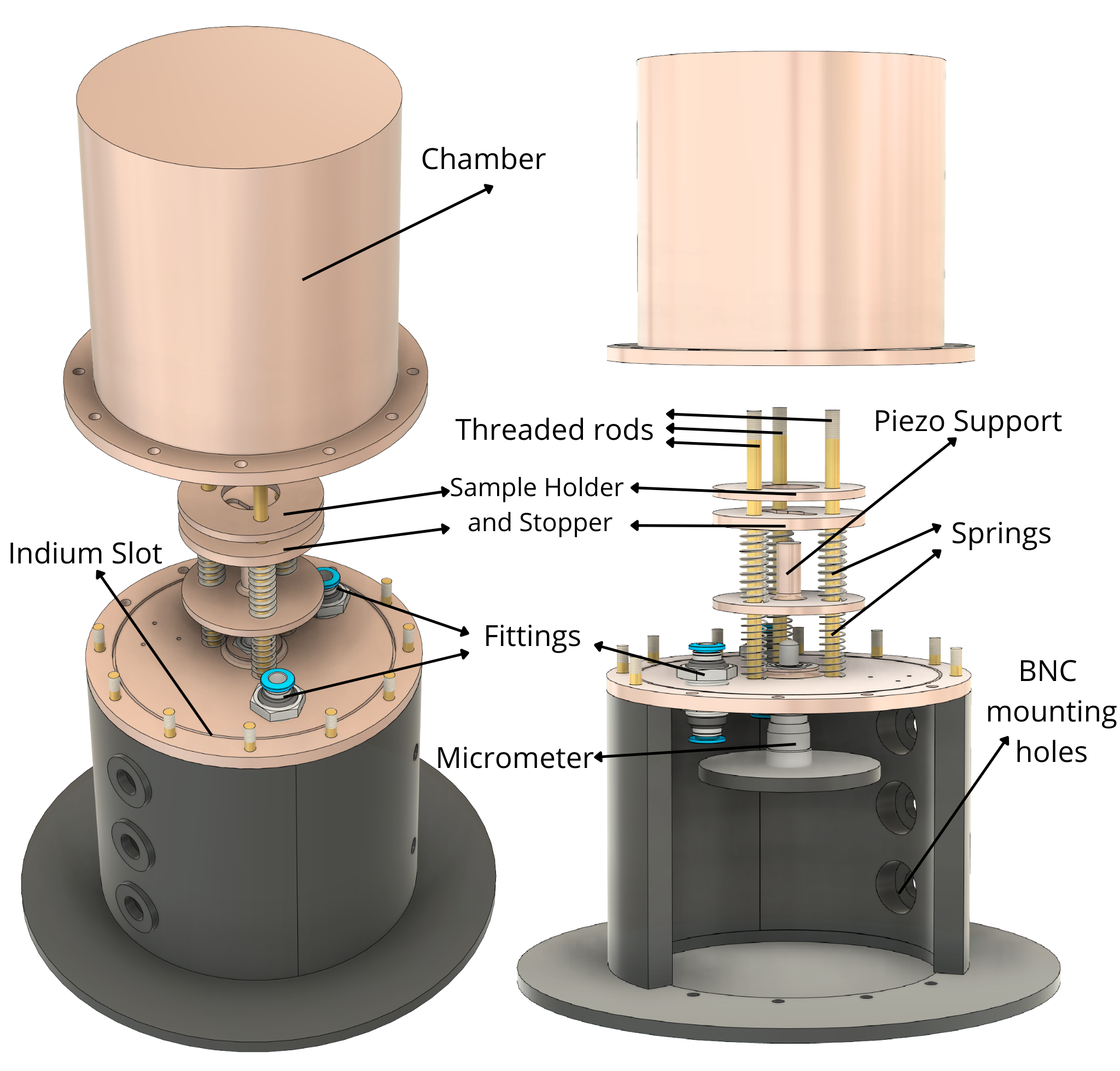}
    \end{subfigure}

    \vspace{0.4cm} 

    \begin{subfigure}{0.6\linewidth}
        \caption{} \label{fig:setupMCbJb}
        \includegraphics[width=\linewidth]{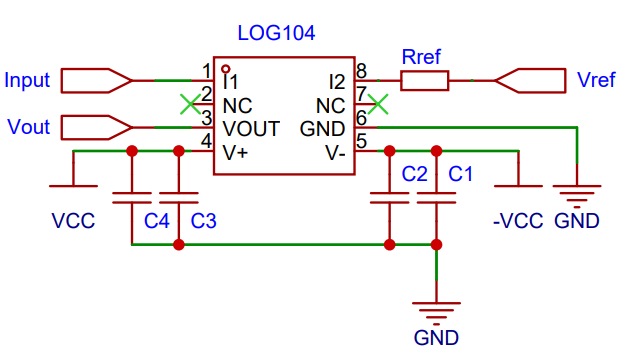}
    \end{subfigure}
    \hfill
    \begin{subfigure}{0.31\linewidth}
        \caption{} \label{fig:setupMCbJc}
        \includegraphics[width=\linewidth]{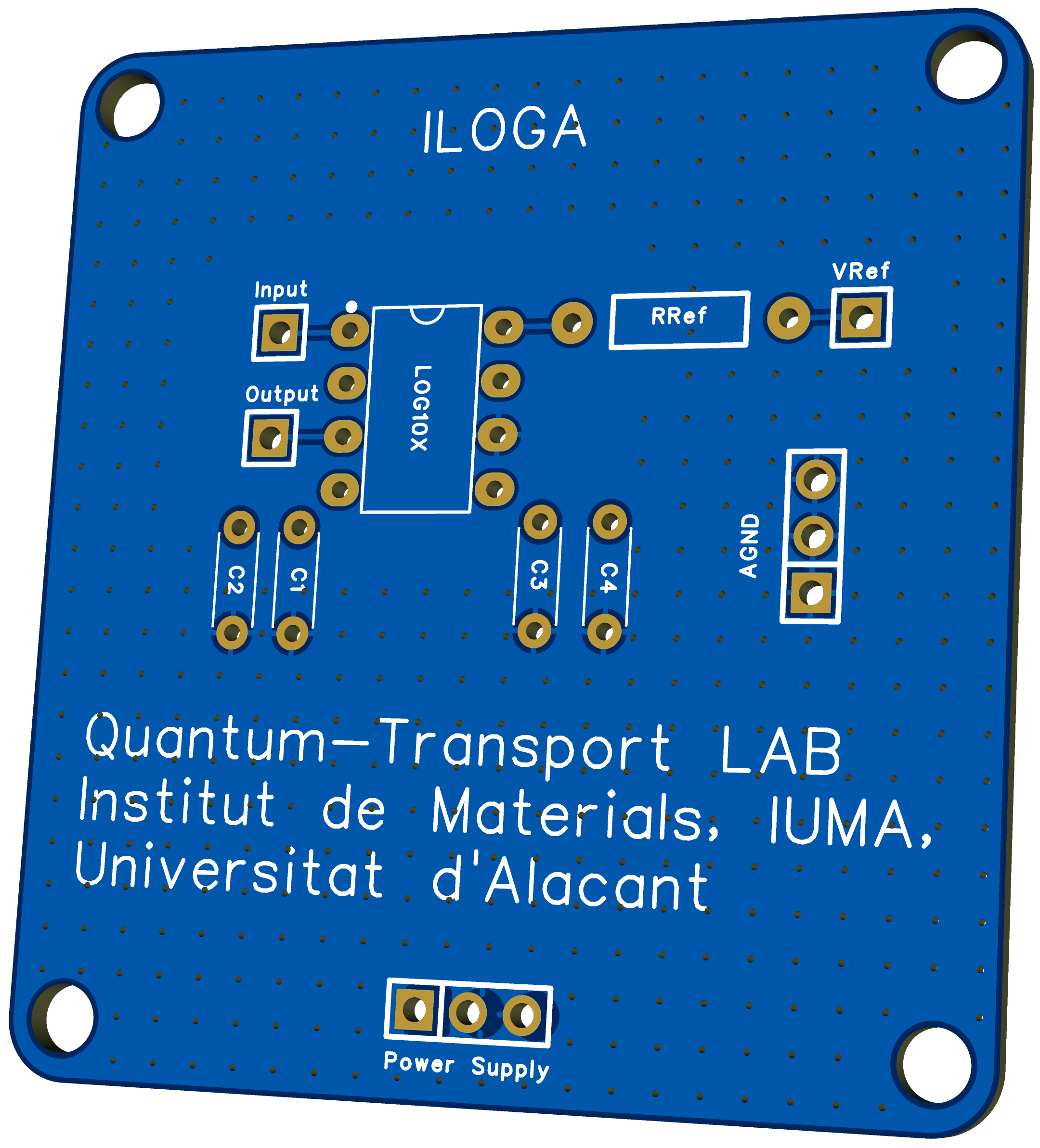}
    \end{subfigure}

    \captionsetup{justification=justified, singlelinecheck=false}
    \caption{\justifying Overview of the experimental hardware and electronics. (a) Detailed CAD rendering of the custom MCBJ setup, highlighting the mechanical architecture designed to operate within the 3D-printed vacuum platform. (b) Circuit schematic of the custom ILOGA system based on the LOG104 integrated circuit, highlighting the $R_{\mathrm{ref}}$ and $V_{\mathrm{ref}}$ configuration. (c) Illustration of the custom 6-layer PCB design engineered for ultra-low noise data acquisition.}
\end{figure}

Within this controlled environment lies the break junction mechanism itself, whose two primary views are presented in Fig.~\ref{fig:setupMCbJa}. The system is designed almost entirely using 3D-printed parts, where the use of non-printed components was strictly limited to the piezoelectric actuator, the micrometer screw, and essential metallic hardware such as guiding rods and springs. In this mechanical architecture, the micrometer screw is responsible for the coarse macroscopic approach, smoothly pushing the piezoelectric element toward the sample to execute the initial mechanical rupture of the metallic wire. Simultaneously, a dedicated spring mechanism provides continuous tension, ensuring the high mechanical stability of both the sample and the sample holder against the guiding rods, thereby minimizing the impact of external vibrations.

\subsection{Logarithmic Amplifier for Quantum Transport} \label{sec:log_amplifier}
We developed a custom $I$--$V$  logarithmic amplifier based on the LOG104 \cite{DatasheetLog104} integrated circuit to overcome the limitations of standard linear equipment. As demonstrated in Ref.~\cite{escorza2026}, linear architectures fail to resolve the extremely low conductance values required for tunneling characterization. 
The operational principle of our design relies on the internal comparison between the transport current from the atomic junction ($I_j$) and a stable reference current ($I_{\mathrm{Ref}}$), entering via Pin 1 and Pin 8, respectively (see Fig.~\ref{fig:setupMCbJb}). This reference current is precisely established by a reference voltage ($V_{\mathrm{Ref}}$) and a reference resistance ($R_{\mathrm{Ref}}$) connected to the device.
This architecture enables the seamless compression of the signal into a manageable voltage output ($V_{\mathrm{out}}$), which is subsequently converted into $G_0$ units following the normalization procedure detailed by Escorza \textit{et al.} \cite{escorza2026}:

\begin{equation}
G = \frac{12\,906}{R_\text{ref}} \cdot \frac{V_\text{ref}}{V_\text{bias}} \cdot 10^{2V_\text{out}} \quad [G_0].
\label{eq:transfer}
\end{equation}

Where $V_{\mathrm{bias}}$ is the constant voltage applied across the metallic junction, which was set to $100$~mV for all experiments in this work. Crucially, this seamless expansion of the dynamic range captures conductance values down to the deep tunneling regime ($\sim 10^{-4}~G_0$) without data loss. This capability enables the extraction of $\phi$ from the tunneling regime in both rupture and formation trace. Notably, while formation dynamics remain steady, the rupture process is inherently more abrupt due to the elastic snap-back of the electrodes upon breaking.

Figure~\ref{fig:setupMCbJb} illustrates the electronic schematic and connections, highlighting the comparison between the current from the atomic junction, $I_j$ (Pin 1), and the stable reference current, $I_{\mathrm{ref}}$ (Pin 8). The custom 6-layer PCB layout is displayed in Fig.~\ref{fig:setupMCbJc}, which incorporates strategic ground planes and electromagnetic shielding to ensure ultra-low noise data acquisition (further details are provided in the Supplementary Material).

To demonstrate the economic feasibility of this platform, a detailed cost breakdown is provided in Table~\ref{tab:costs} (Supplementary Material), showing that the complete MCBJ and custom $I$--$V$ amplification system remain highly accessible at approximately 460 \euro{}.

\section{Results and Discussion}

After validating the 3D-printed chamber and MCBJ mechanism, we measured electronic transport using Au and Cu electrodes across three environments. First, we tested under room conditions to establish a baseline for environmental contamination and oxidation. Second, we applied high vacuum to prevent the rapid oxidation of Cu and ensure clean quantum transport signatures. Finally, we immersed the junctions in  glycerol to evaluate stability in a viscous liquid. While the high-vacuum environment successfully protected the Cu electrodes, environmental conditions profoundly impacted the mechanical stability. Although the direct coupling of the turbomolecular pump introduces mechanical noise that slightly reduces the length of the conductance plateaus, our custom logarithmic amplifier successfully captures the fundamental quantum transport signatures (such as the sharp $1G_0$ peak) without data loss. Future iterations of the platform could incorporate mechanical decoupling to further enhance atomic stability, but the current structural rigidity is thoroughly sufficient to resolve the tunneling regime.

\begin{table}[h!]
\centering
\caption{\justifying Summary of the rupture trace statistics for Au and Cu atomic contacts. Valid traces are calculated based on the success rate.}

\label{tab:trace_statistics}
\resizebox{\columnwidth}{!}{%
\begin{tabular}{@{}llccc@{}}
\toprule
\textbf{Environment} & \textbf{Metal} & \textbf{Total Traces} & \textbf{Valid Traces} & \textbf{Success Rate} \\ \midrule
\textbf{Room Conditions} & Au & 3000 & 2997 & $\approx$99.9\% \\
 & Cu & 1000 & 4 & $\approx$0.04\% \\ \midrule
\textbf{HV in 3D Chamber} & Au & 3600 & 3593 & $\approx$99.8\% \\
 & Cu & 2000 & 1984 & $\approx$99.2\% \\ \midrule
\textbf{Glycerol} & Au & 7600 & 7592 & $\approx$99.9\% \\
 & Cu & 575 & 540 & $\approx$93.9\% \\ \bottomrule
\end{tabular}%
}
\end{table}

\begin{figure*}[htp]
    \centering
    \captionsetup[subfigure]{position=top, justification=raggedright, singlelinecheck=false, labelfont=normalsize}

    \begin{subfigure}{0.45\textwidth}
        \caption{} \label{fig:gold_traces}
        \includegraphics[width=\linewidth]{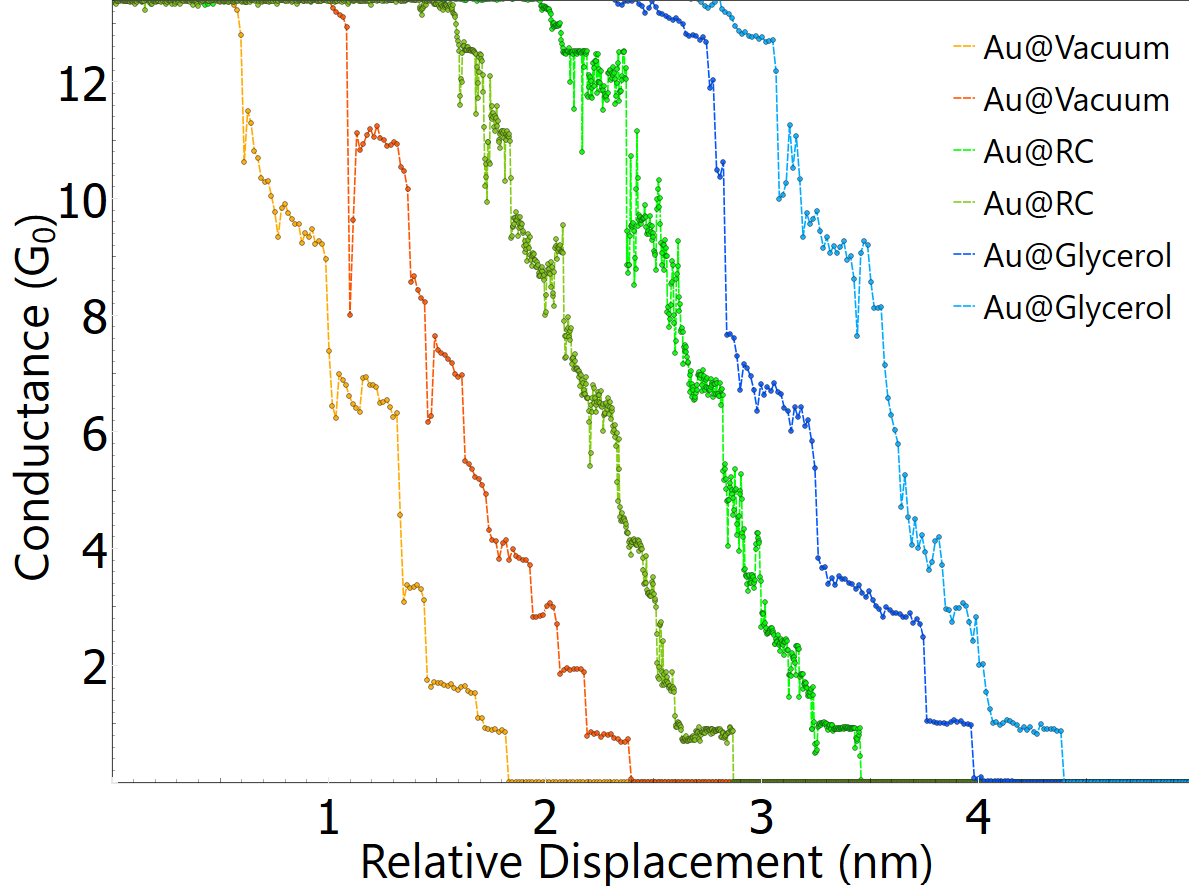}
    \end{subfigure}
    \hfill
    \begin{subfigure}{0.45\textwidth}
        \caption{} \label{fig:copper_traces}
        \includegraphics[width=\linewidth]{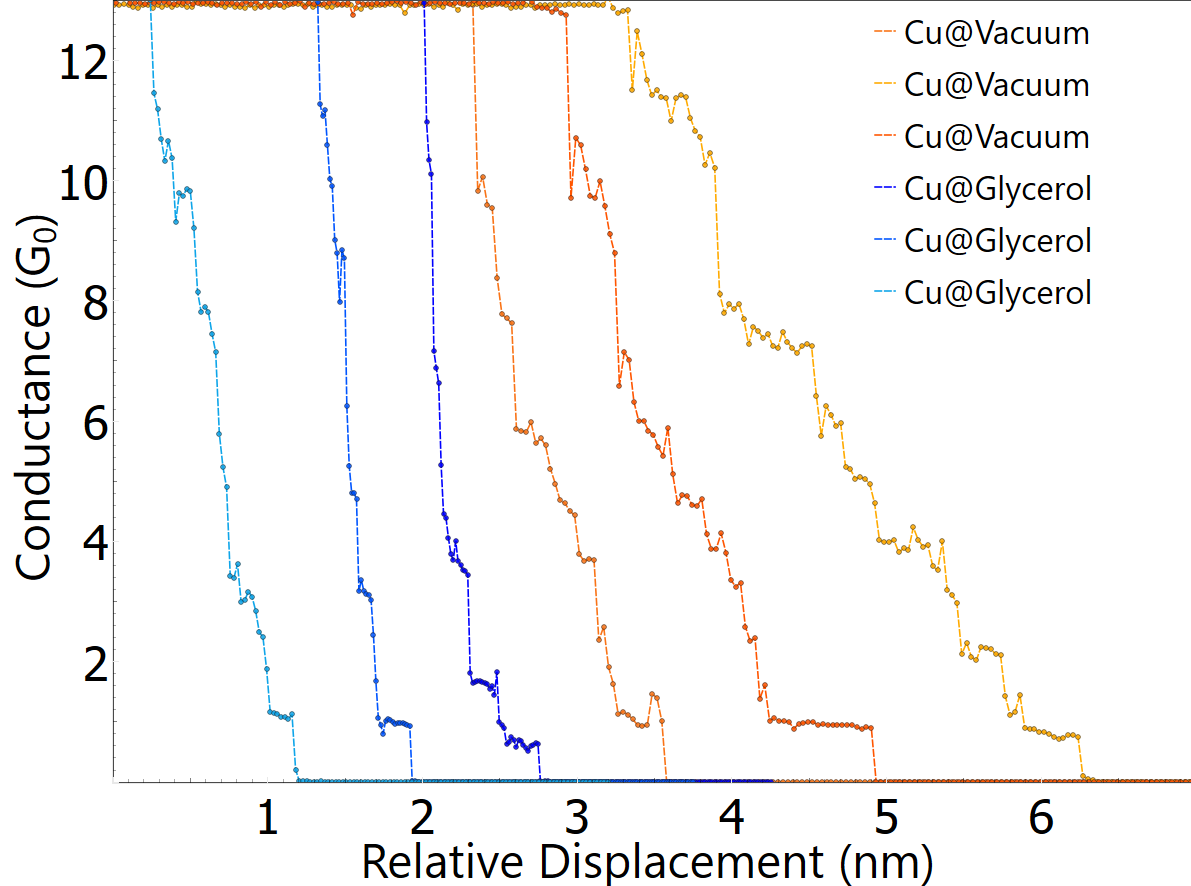}
    \end{subfigure}

    \vspace{0.1cm} 

    \begin{subfigure}{0.45\textwidth}
        \caption{} \label{fig:gold_hist}
        \includegraphics[width=\linewidth]{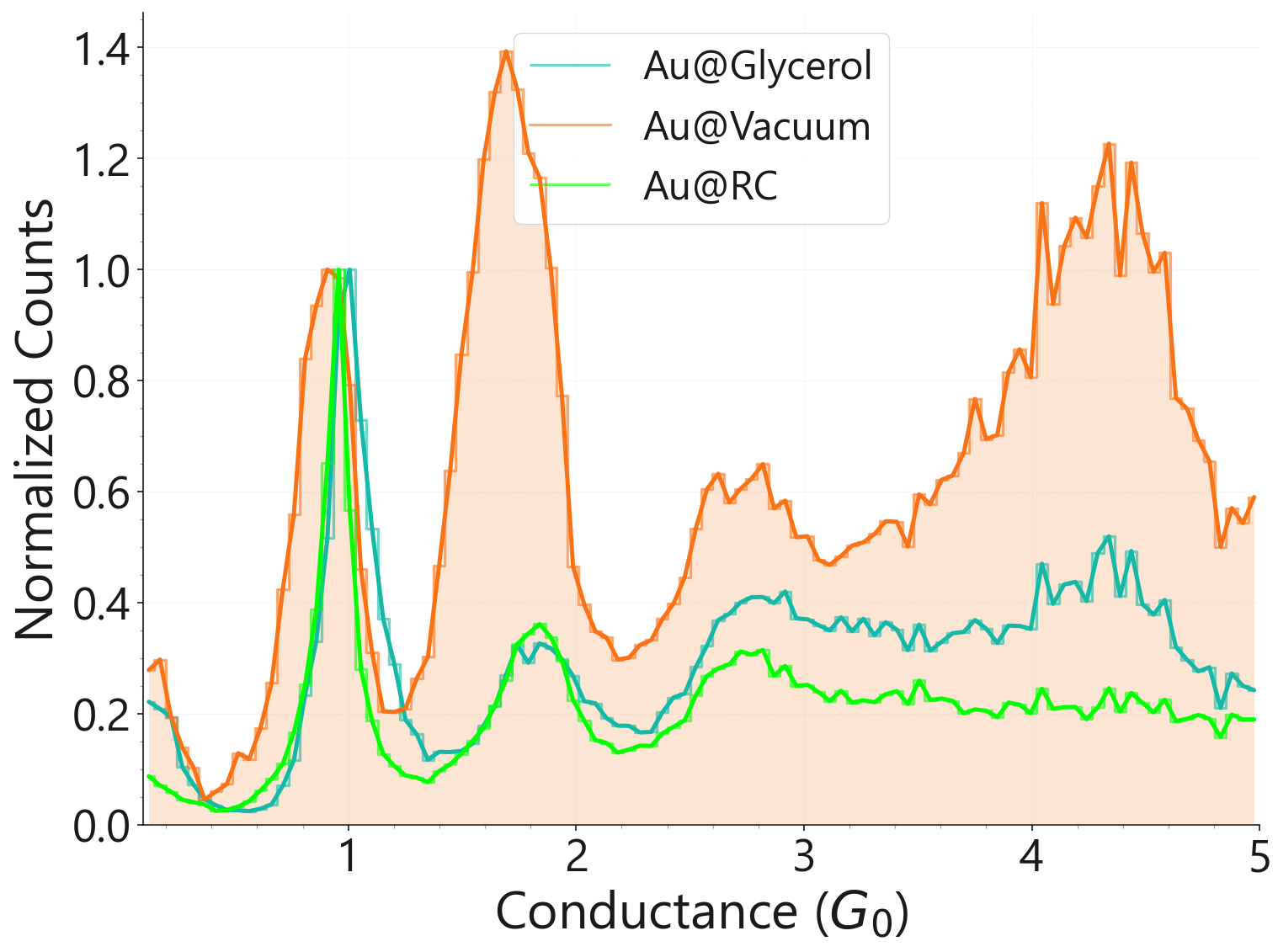}
    \end{subfigure}
    \hfill
    \begin{subfigure}{0.45\textwidth}
        \caption{} \label{fig:copper_hist}
        \includegraphics[width=\linewidth]{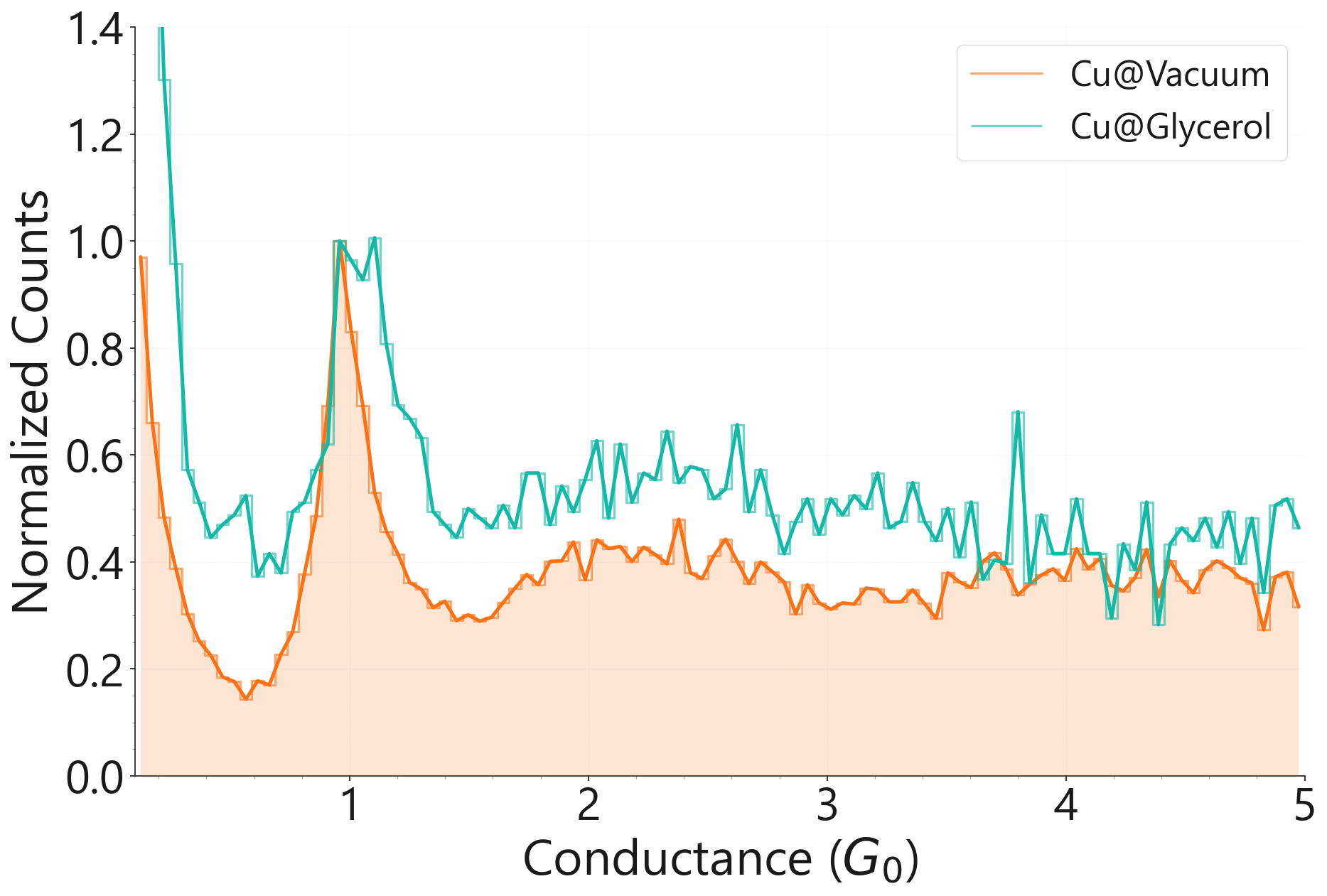}
    \end{subfigure}
 \captionsetup[subfigure]{position=top, justification=justified, singlelinecheck=false, labelfont=small}
 \captionsetup{justification=justified, singlelinecheck=false}
    \caption{\justifying Rupture traces and conductance histograms of Au and Cu in different environments. (a) Individual rupture traces of Au obtained in vacuum, glycerol, and ambient conditions. (b) Individual rupture traces of Cu in vacuum and glycerol. (c) Conductance histograms of Au normalized to $1G_0$ for the three different environments. (d) Normalized conductance histograms comparing Cu in vacuum and glycerol.}
    \label{fig:GolResults}
\end{figure*}

In this study, our statistical analysis is based exclusively on rupture traces. Although the tunneling decay region in rupture traces is typically shorter than during the formation process, they are generally much less susceptible to contamination and atomic rearrangements, making them the preferred and most reliable standard for molecular electronics. Table \ref{tab:trace_statistics} summarizes the rupture trace statistics for both Au and Cu atomic contacts across the three tested environments: Room Conditions (RC), High Vacuum (HV) inside the 3D-printed chamber, and immersed in glycerol. The table is organized to explicitly detail the platform's performance. The first two columns indicate the experimental environment and the respective metal. The third column displays the total number of traces attempted and recorded by the acquisition software. The fourth column presents the number of valid traces; this number is obtained by carefully filtering out traces where the contact did not break completely,  signals remained saturated, or simply dropped to the baseline noise level of the I-V converter. Finally, the fifth column calculates the success rate of the measurement. As clearly demonstrated by this success rate, acquiring valid Cu traces under room conditions is practically impossible ($\approx 0.04\%$) due to rapid surface oxidation. However, both the 3D-printed HV chamber and the glycerol immersion act as highly effective inert media, preventing degradation and dramatically restoring the viability of Cu measurements to over $93\%$. Conversely, Au atomic contacts prove to be highly resilient, maintaining a nearly perfect success rate across all environments.

Following the statistical analysis, Fig~\ref{fig:GolResults} presents the individual rupture traces and their corresponding conductance histograms for Au (left panels, (a) and (c)) and Cu (right panels, (b) and (d)). In  Fig~\ref{fig:GolResults}(a), typical Au rupture traces are displayed using a consistent color code: green tones for RC, brown/ochre for HV, and blue for immersion in  glycerol. Applying this same color scheme, panel (b) displays the Cu traces in HV and glycerol. Notably, RC traces for Cu are omitted entirely, as severe atmospheric oxidation prevents the formation of any measurable atomic contacts. Panels (c) and (d) display the conductance histograms constructed from these traces. For Au (panel (c)), the characteristic $1G_0$ peak is robustly present across all environments, and the overall transport behavior above $0.2G_0$ remains virtually identical regardless of the medium. The slight variations in the HV histogram, here the $1G_0$ peak appears more pronounced, are attributed to the mechanical instability introduced by the vacuum pump's vibrations. As seen in the individual traces (panel a), this vibrational noise causes the conductance plateaus to be shorter and less stable, slightly altering the relative peak dimensions. Conversely, the Cu histograms (panel d) reveal distinct environmental interactions. Under HV conditions, Cu exhibits a sharp, narrow $1G_0$ peak that closely resembles the behavior of Au and is consistent with cryogenic (4.2~K) measurements reported in the literature \cite{phdYanson,Role18}. However, when immersed in glycerol, the $1G_0$ peak broadens significantly, indicating that while the liquid successfully prevents macroscopic oxidation, the highly reactive Cu still interacts subtly with the dielectric medium. Ultimately, these results confirm that both the 3D-printed high-vacuum chamber and glycerol immersion serve as effective protective environments to resolve the quantum transport signatures of reactive metals.

\begin{figure*}[]
    \centering
    \captionsetup[subfigure]{position=top, justification=raggedright, singlelinecheck=false, labelfont=normalsize}

    \begin{subfigure}{0.32\textwidth}
        \caption{} \label{fig:phyRC}
        \includegraphics[width=\linewidth]{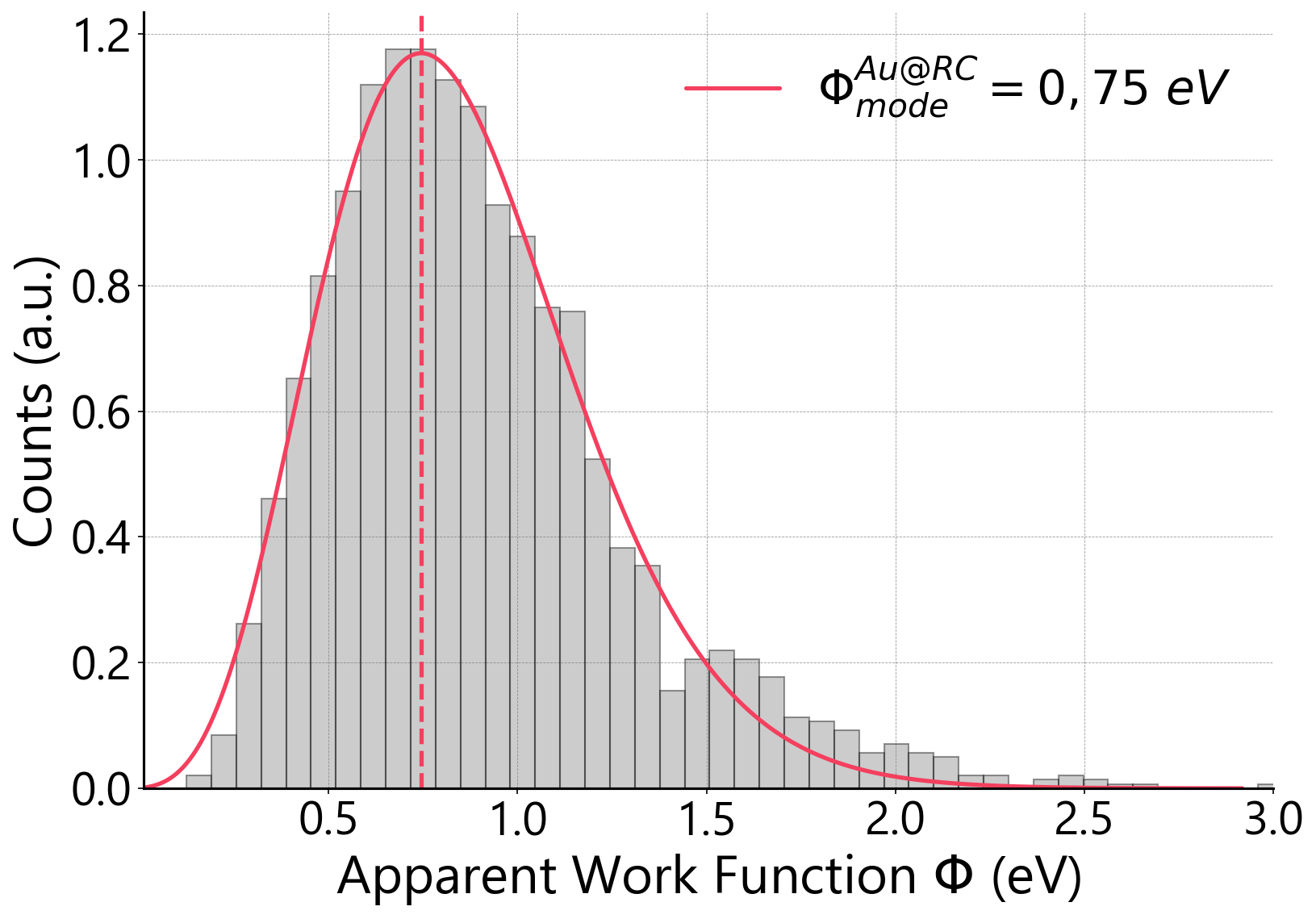}
    \end{subfigure}\hfill
    \begin{subfigure}{0.32\textwidth}
        \caption{} \label{fig:phyvac}
        \includegraphics[width=\linewidth]{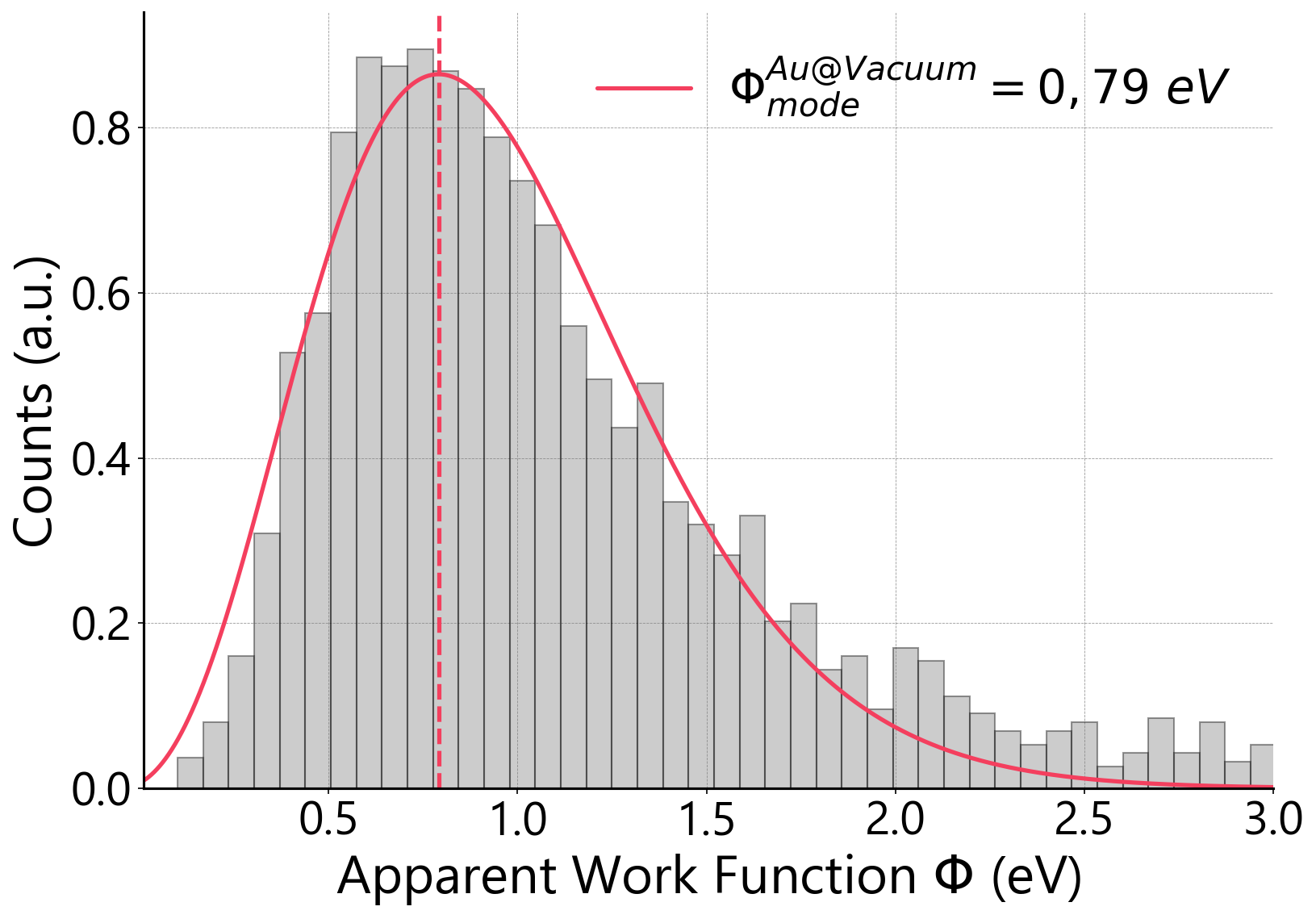}
    \end{subfigure}\hfill
    \begin{subfigure}{0.32\textwidth}
        \caption{} \label{fig:phygly}
          \includegraphics[width=\linewidth]{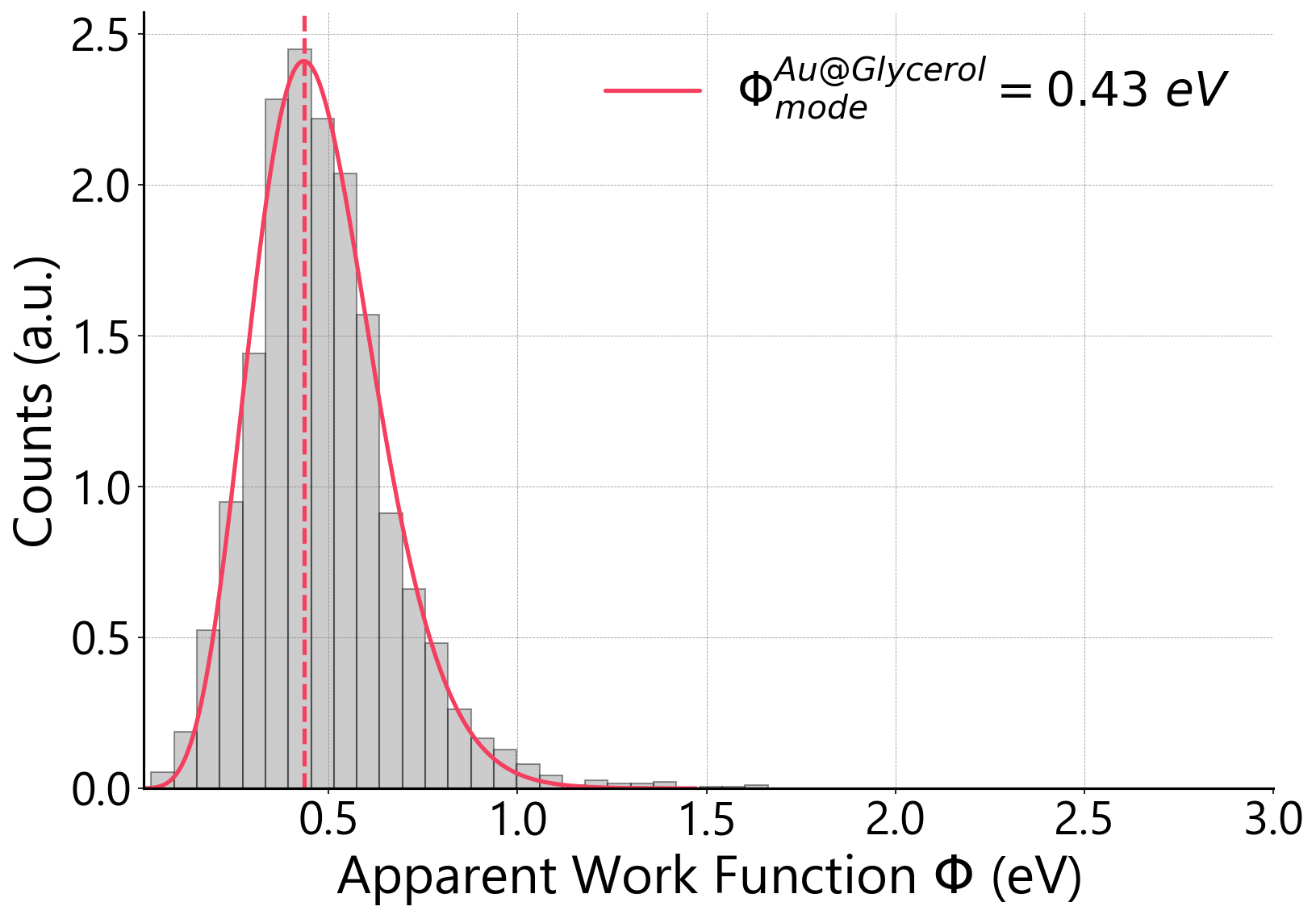}
        
    \end{subfigure}
    
    \captionsetup[subfigure]{position=top, justification=justified, singlelinecheck=false, labelfont=small}
    \captionsetup{justification=justified, singlelinecheck=false}
    
\caption{\justifying Apparent work function histograms of Au across different environments: (a) under room conditions, (b) in high vacuum inside the 3D-printed chamber, and (c) immersed in glycerol.}
    \label{fig:WF}
\end{figure*}

Having validated the platform's capability to preserve and resolve atomic contacts, we now turn our attention to the tunneling regime to evaluate the surface energetics across these environments. We extract $\phi$ exclusively from formation traces, where the tunneling regime is more detectable. By taking the natural logarithm of \cref{eq:tunneling}, the decay is linearized with a slope given by $S = \frac{2\sqrt{2m_e\phi}}{\hbar}$. From this relation, the work function is easily isolated as $\phi = \frac{\hbar^2 S^2}{8m_e}$. A least-squares linear regression is applied to the well-defined tunneling region to extract this slope $S$. Further details regarding this mathematical conversion and error propagation are provided in the Supplementary Material, where, as an illustrative example, the fit of \cref{fig:tunnelingtrace} yields an apparent work function of $1.31 \pm 0.02$~eV. However, since a single trace only provides an individual measurement, understanding the overall statistical behavior of the work function requires extracting $\phi$ from thousands of traces to construct representative histograms.

As illustrated in Fig. \ref{fig:XiGau}, the histogram is rigorously modeled by a non-central chi-square ($\chi^2$) distribution. This analytical choice arises from a well-established statistical property:\cite{Taylor1997,Hughes2010} squaring a normally distributed random variable fundamentally alters its profile into this exact distribution (see the slope histogram in the inset of Fig. \ref{fig:XiGau}).  It is important to emphasize the physical origin of this statistical behavior. The Gaussian distribution of the tunneling slopes stems from the inherent structural variability of the atomic contact during its formation and rupture. Following the mechanical snap-back\cite{Trouwborst}, the electrodes undergo random atomic reconfigurations, leading to diverse crystallographic geometries at the apex. Furthermore, phenomena such as the Smoluchowski \cite{Smoluchowski,WFAnisotropy}smoothing effect create local variations in the surface dipole depending on this exact atomic arrangement. These physical variations directly modulate the tunneling decay constant (the slope), yielding a normal distribution. Consequently, the non-central chi-square profile of the apparent work function is strictly a mathematical manifestation of its quadratic dependence on this normally distributed slope ($\phi \propto S^2$), rather than a distribution with an independent physical origin. 

Since the work function depends quadratically on this normally distributed slope, its probability density is mathematically defined as:

\begin{equation}
P(x; \lambda) = \frac{1}{2} \exp\left(-\frac{x + \lambda}{2}\right) \left(\frac{x}{\lambda}\right)^{-1/4} I_{-1/2}(\sqrt{\lambda x})
\label{eq:chi_square}
\end{equation}

where $P(x)$ is the probability density function of the scaled variable $x$, $\lambda$ is the non-centrality parameter, and $I_{-1/2}$ represents the modified Bessel function of the first kind. Because the apparent work function is derived from squaring a single variable (the tunneling slope), the system possesses exactly one degree of freedom ($k=1$). Consequently, the experimental histograms were fitted using the 1-degree-of-freedom non-central chi-square model (\texttt{scipy.stats.ncx2}) from the \texttt{SciPy} library~\cite{Scipy} within a custom \texttt{Python}~\cite{python3} script. Among other variables, the fitting algorithm returns the non-centrality parameter ($\lambda$) and the scale factor (sf). From these, we numerically extract the most statistically and physically relevant values for our analysis: the mode of $\phi$ and the standard deviation ($\sigma$). For our $1$-degree-of-freedom system, this dispersion is analytically computed as $\sigma = \text{sf} \times \sqrt{2(1 + 2\lambda)}$. This establishes the corresponding $1\sigma$ range containing approximately $68\%$ of the data.

Figure~\ref{fig:WF} presents the distributions of the $\phi$ obtained from the formation traces across the three distinct environments: (a) room conditions (RC), (b) high vacuum, and (c) glycerol. After computing the tunneling slopes from thousands of valid traces, we constructed the corresponding $\phi$ histograms. In all three cases, the experimental data (gray bins) exhibit a characteristic positive skewness (right-tail asymmetry); therefore, a non-central chi-square $\chi^2$ function (solid red curve) was rigorously fitted to each histogram. Additionally, the mode extracted from this fit is marked by a vertical red dashed line. Table~\ref{tab:work_function_summary} presents the apparent work function modes and standard deviations ($\sigma$) calculated from the $\chi^2$ fits for each environment.

\begin{table}[htpb]
\caption{\justifying Summary of the $\chi^2$ fit results from Fig.~\ref{fig:WF} for Au in different environments (first column). The second and third columns present the mode ($\phi$) and standard deviation ($\sigma$), respectively, while the fourth and fifth columns report the optimized non-centrality parameter ($\lambda$) and scale factor (sf).}
\centering
\begin{tabular}{lcccc}
\hline
\hline
\textbf{Environment} & \textbf{$\phi_{\text{mode}}$ (eV)} & \textbf{$\sigma$ (eV)} & \textbf{$\lambda$} & \textbf{sf} \\
\hline
Room Conditions & 0.75 & 0.35 & 22.6 & 0.04 \\ 
High Vacuum     & 0.79 & 0.44 & 16.9 & 0.05 \\
Glycerol        & 0.43 & 0.16 & 32.5 & 0.01 \\
\hline
\hline
\end{tabular}
\label{tab:work_function_summary}
\end{table}

Across all environments, the measured $\phi$ values are significantly lower than the $\sim 5.3$~eV expected for pristine gold. This reduction is primarily driven by ubiquitous atmospheric adsorbates (such as $\text{O}_2$ and $\text{H}_2\text{O}$). As recently reviewed by Nguyen and Lenfant~\cite{Nguyen2026}, these molecules create a strong surface dipole that shifts the vacuum level and drastically lowers the apparent work function. Moreover, when immersed in glycerol, $\phi$ drops even further. In this environment, the tunneling process occurs directly through a dielectric liquid rather than air or vacuum; consequently, the physical presence of these organic molecules within the atomic gap effectively lowers the electron tunneling barrier. Furthermore, a closer inspection of the $\chi^2$ distributions provides additional insight into the high vacuum measurements. Although the mode of the high vacuum distribution remains centered at a low value (comparable to room conditions), its right tail extends noticeably further towards higher values. While this pronounced tail indicates that a small fraction of the junction ruptures approach the $\sim 5.3$~eV expected for pristine gold, the persistently low mode confirms that the vacuum environment alone is insufficient to fully desorb the initially physisorbed molecules. Consequently, even when the nanocontact breaks under vacuum, these atmospheric adsorbates remain predominantly bound to the electrode surfaces, continuing to dictate the most probable electron tunneling barrier.

Finally, while these relative environmental trends are completely robust, it must be noted that the absolute $\phi$ values are intrinsically sensitive to the experimental setup. As detailed in the Supplementary Material (see figure \ref{fig:WFSM}), a standard $10\%$ experimental uncertainty in the piezoelectric displacement calibration propagates quadratically, introducing an approximate $20\%$ error in the absolute work function calculation only for the case of gold.

\section{Conclusions}
In summary, we have designed, fabricated, and validated a low-cost, 3D-printed MCBJ platform that enables high-resolution quantum transport measurements of reactive metals at room temperature. We have demonstrated that standard PLA filament can sustain stable high-vacuum conditions ($1.4 \times 10^{-4}$ mbar) and withstand a 1 atm pressure differential, providing a robust and sustainable alternative to expensive commercial vacuum systems. Our results show that both the high-vacuum environment and glycerol immersion are highly effective at preventing the rapid oxidation of copper, increasing the measurement success rate.

Furthermore, leveraging our custom logarithmic amplifier, we systematically extracted the apparent work function $\phi$ from thousands of individual formation traces using gold as a robust benchmark. We demonstrated that the statistical distribution of $\phi$ strictly follows a non-central chi-square profile. This arises mathematically from normally distributed tunneling slopes driven by random atomic reconfigurations and diverse apex geometries~\cite{Role18}. Crucially, as established in scanning tunneling microscopy~\cite{Noei2018}, this atomic-scale topography dictates the lateral confinement of the electron wave function, introducing a variable kinetic-energy contribution that inherently scatters the measured barrier height. Across ambient air, high vacuum, and anhydrous glycerol, the most probable work functions (modes) were $0.75$~eV, $0.79$~eV, and $0.43$~eV, respectively. Although fundamentally lower than theoretical expectations for pristine metals, this reduction is comprehensively explained by the combined effects of local surface dipoles from intrinsic roughness~\cite{Smoluchowski,WFAnisotropy}, the aforementioned kinetic-energy contributions at the atomic apex~\cite{Noei2018}, atmospheric adsorbates~\cite{VanWee2005}, and dielectric liquid interactions~\cite{Nguyen2026}.

Even with these environmentally altered surface energetics, the ability to clearly resolve the $1G_0$ conductance quantum and the subsequent exponential tunneling decay represents a significant milestone for accessible nanotechnology. Despite these environmental interactions, Cu remains a superior electrode material compared to Au due to its optimal Fermi level alignment with molecular HOMO levels and enhanced $d$-orbital coupling. By drastically lowering the economic and technical barriers of traditional fabrication, this platform not only democratizes the study of copper-based molecular junctions but also opens the door to the systematic characterization of a wide catalog of reactive materials, such as Platinum (Pt), Nickel (Ni), Aluminum (Al), and Palladium (Pd). Ultimately, this methodology establishes 3D printing technology as a viable path for developing precision scientific instrumentation for next-generation molecular electronics.

\begin{acknowledgments}
The authors gratefully acknowledge financial support from the Generalitat Valenciana (CIDEXG/2022/45) and the Spanish Government through  PID2023-146660OB-I00, received funding from MICIU/AEI/10.13039/501100011033, and the European Regional Development Fund (ERDF/EU). The authors also wish to thank Dr. Werner Bramer from Yachay Tech University and Prof. Carlos Untiedt from Alicante University for fruitful discussions.
\end{acknowledgments}

\clearpage
\section{REFERENCES}
 \bibliography{Refe3dPrintInner}

\newpage
\section*{Supplementary Material}

\subsection{Logarithmic Amplifier: Power Supply and Shielding Details}
To ensure the optimal performance of the custom logarithmic amplifier, the power supply rails require strict filtering. To mitigate high-frequency noise and prevent detrimental oscillations, these rails are heavily decoupled by a network of bypass capacitors (C1--C4) connected to the common ground (Pin 6). Additionally, pins 2 and 7 of the LOG104 are kept as non-connected (NC) to prevent parasitic interference with the highly sensitive signal traces.

 \subsection{MCBJ costs}
The table \ref{tab:costs} accounts for all essential components, including the 3D-printed vacuum chamber, the integrated MCBJ assembly, and the custom logarithmic amplification system. By disclosing these costs alongside our open-source design files, we aim to provide a transparent reference for researchers seeking to implement high-performance, cost-effective quantum transport experiments.

\setcounter{table}{0}
\renewcommand{\thetable}{S\arabic{table}}
\setcounter{figure}{0}
\renewcommand{\thefigure}{S\arabic{figure}}
\setcounter{equation}{0}
\renewcommand{\theequation}{S\arabic{equation}}

\begin{table}[htp]
\captionsetup[subfigure]{position=top, justification=justified, singlelinecheck=false, labelfont=small}
\caption{\justifying Cost breakdown of the custom MCBJ setup components.}
\centering
\label{tab:costs}
\renewcommand{\arraystretch}{1.2}
\resizebox{\columnwidth}{!}{
\begin{tabular}{llc}
\toprule
\textbf{Component} & \textbf{Description} & \textbf{Estimated Cost (\euro{})} \\ 
\midrule
Vacuum Chamber & 3D-printed (Resin/Filament) & $\sim30$--$50$ \\ 
ILOGA Amplifier & Custom PCB (6-layer) + Components & $\sim25$--$50$ \\ 
Mechanical Control & Micrometer screw and springs & $\sim35$--$60$ \\ 
Piezoelectric Actuator & Low-cost stack/plate piezo & $\sim100$--$300$ \\ 
\midrule
\textbf{Total System} & \textbf{Estimated total investment} & \textbf{$\sim190$--$460$} \\ 
\bottomrule
\end{tabular}%
}
\end{table}

\subsection{Determination of the Apparent Work Function}

The $\phi$ is extracted from the exponential tunneling regime during the contact formation process. By taking the natural logarithm of \cref{eq:tunneling} from the main text, the conductance decay is linearized as follows:
\begin{equation}
    \ln(G) = \ln(G_i) - \left(\frac{2\sqrt{2m_e\phi}}{\hbar}\right)z = \ln(G_i) - S z
\end{equation}
To determine $\phi$, a least-squares linear fit is applied directly to the experimental data within the tunneling regime. The resulting slope $S$ from this fit corresponds to the theoretical prefactor $2\sqrt{2m_e\phi}/\hbar$. By isolating the work function, we obtain:
\begin{equation}
    \phi = \frac{\hbar^2}{8m_e} S^2
    \label{eq:fiisolated}
\end{equation}

\begin{figure}[htp]
    \centering
    \includegraphics[width=\linewidth]{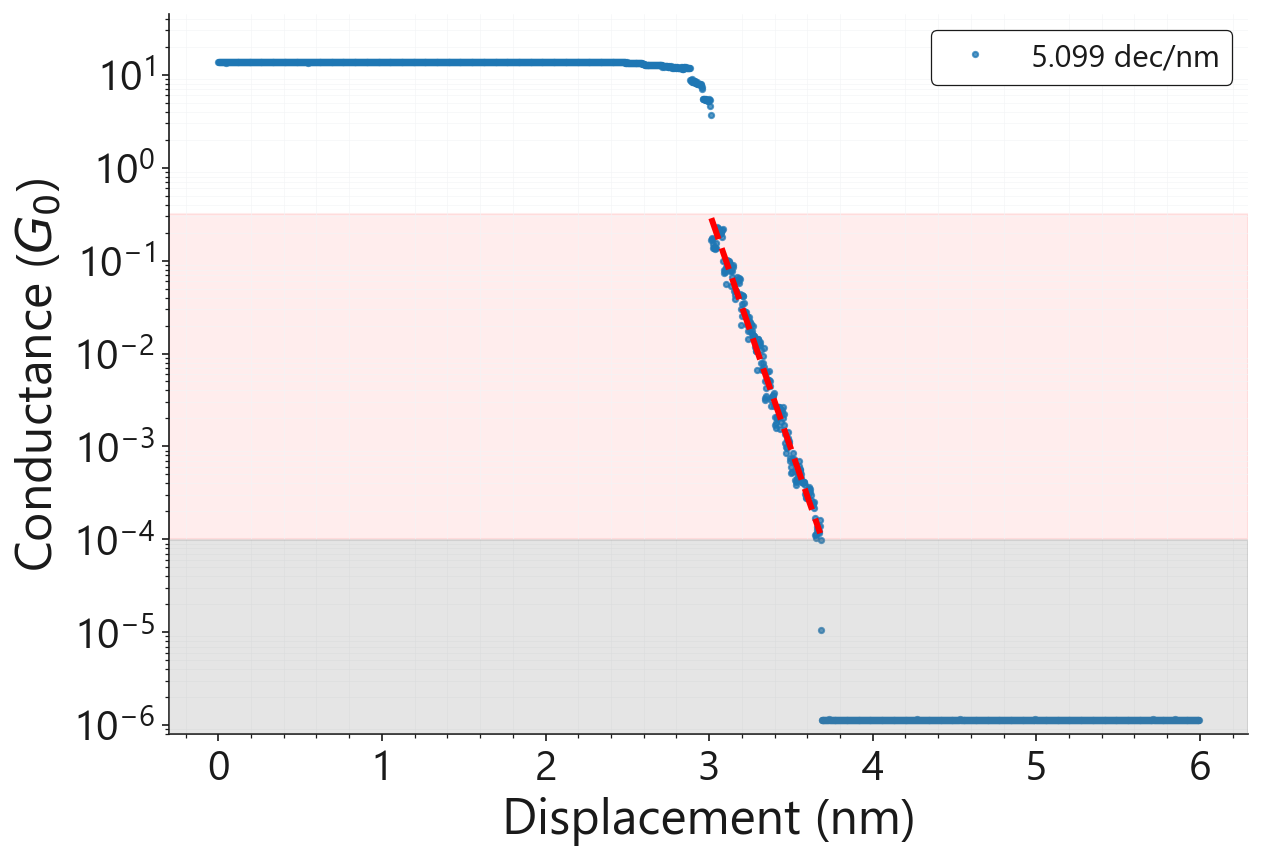}
    \caption{\justifying Representative formation trace. The blue points represent the experimental measurements, while the continuous red line corresponds to the least-squares linear fit applied in the tunneling regime. The pink highlighted
area shows the valid measurement range; in contrast,
the grey area indicates the unreliable region caused by
the logarithmic amplifier architecture (for more details,
see\cite{escorza2026}). The legend displays the extracted slope expressed in decades per nanometer.}
    \label{fig:tunnelingtrace}
\end{figure}

As illustrated in \cref{fig:tunnelingtrace}, this methodology allows for the clear differentiation of the metallic (upper), tunneling (middle), and non-validity (lower) regions, yielding a high-quality linear fit ($R^2 = 0.9851$). 

To illustrate the extraction process, we calculate the $\phi$ for the experimental slope of $S_{\text{dec}} = 5.099$~dec/nm shown in the figure. First, we convert the experimental slope from decades per nanometer to the natural logarithm scale ($S$) and express it in inverse \AA ngstr\"oms (\AA$^{-1}$):
\begin{equation}
\begin{split}
    S &= S_{\text{dec}} \ln(10) = 5.099 \times 2.3026 \\
      &\approx 11.741 \text{ nm}^{-1} = 1.174 \text{ \AA}^{-1}
\end{split}
\end{equation}

Next, we evaluate the constant prefactor $\hbar^2 / 8m_e$ from \cref{eq:fiisolated}. To do this explicitly and avoid cumbersome exponents, it is highly convenient to multiply both the numerator and the denominator by the speed of light squared ($c^2$). This allows us to express the term using well-known fundamental physics constants in electron-volts:
\begin{equation}
    \frac{\hbar^2}{8m_e} = \frac{\hbar^2 c^2}{8 m_e c^2} = \frac{(\hbar c)^2}{8 (m_e c^2)}
\end{equation}

We can now substitute the standard values for the electron rest mass energy ($m_e c^2 \approx 511 \times 10^3 \text{ eV}$) and the reduced Planck constant multiplied by the speed of light ($\hbar c \approx 197.327 \text{ eV}\cdot\text{nm}$). Converting the latter into \AA ngstr\"oms ($1 \text{ nm} = 10 \text{ \AA}$) yields $\hbar c \approx 1973.27 \text{ eV}\cdot\text{\AA}$. 

Substituting these values, we obtain the numerical constant:
\begin{equation}
    \frac{(\hbar c)^2}{8 (m_e c^2)} = \frac{(1973.27 \text{ eV}\cdot\text{\AA})^2}{8 \times 510\,999 \text{ eV}} \approx 0.9525 \text{ eV}\cdot\text{\AA}^2
\end{equation}

Finally, by substituting this constant and our previously converted experimental slope ($S = 1.174 \text{ \AA}^{-1}$) back into \cref{eq:fiisolated}, we obtain the final apparent work function:
\begin{equation}
    \phi \approx 0.9525 \cdot (1.174)^2 \approx 1.312 \text{ eV}
\end{equation}

To complete the analysis and rigorously quantify the uncertainty of the measurement, we perform a standard error propagation based on \cref{eq:fiisolated}. Given the quadratic dependence of the work function on the slope ($\phi \propto S^2$), the relative error of $\phi$ is twice the relative error of $S$:
\begin{equation}
    \frac{\Delta \phi}{\phi} = 2 \frac{\Delta S}{S} = 2 \frac{\Delta S_{\text{dec}}}{S_{\text{dec}}}
\end{equation}

For the representative fit shown in the figure, the statistical error of the slope extracted from the least-squares regression is $\Delta S_{\text{dec}} = 0.04$~dec/nm. Applying this uncertainty to our previously calculated value, the absolute error is:
\begin{equation}
    \Delta \phi = 2 \phi \left( \frac{\Delta S_{\text{dec}}}{S_{\text{dec}}} \right) = 2 (1.312 \text{ eV}) \left( \frac{0.04}{5.099} \right) \approx 0.02 \text{ eV}
\end{equation}

Therefore, directly incorporating the fitting error, the final apparent work function extracted from this individual trace is rigorously reported as:
\begin{equation}
    \phi = 1.31 \pm 0.02\text{ eV}
\end{equation}

\subsection{Statistical Analysis of Slopes and Work Functions}

\begin{figure}[htbp]
    \centering
    \vspace{0.5cm} 
    
    \setlength{\fboxsep}{0pt}
    \setlength{\fboxrule}{0.5pt} 
    
    \begin{overpic}[width=\columnwidth, percent]{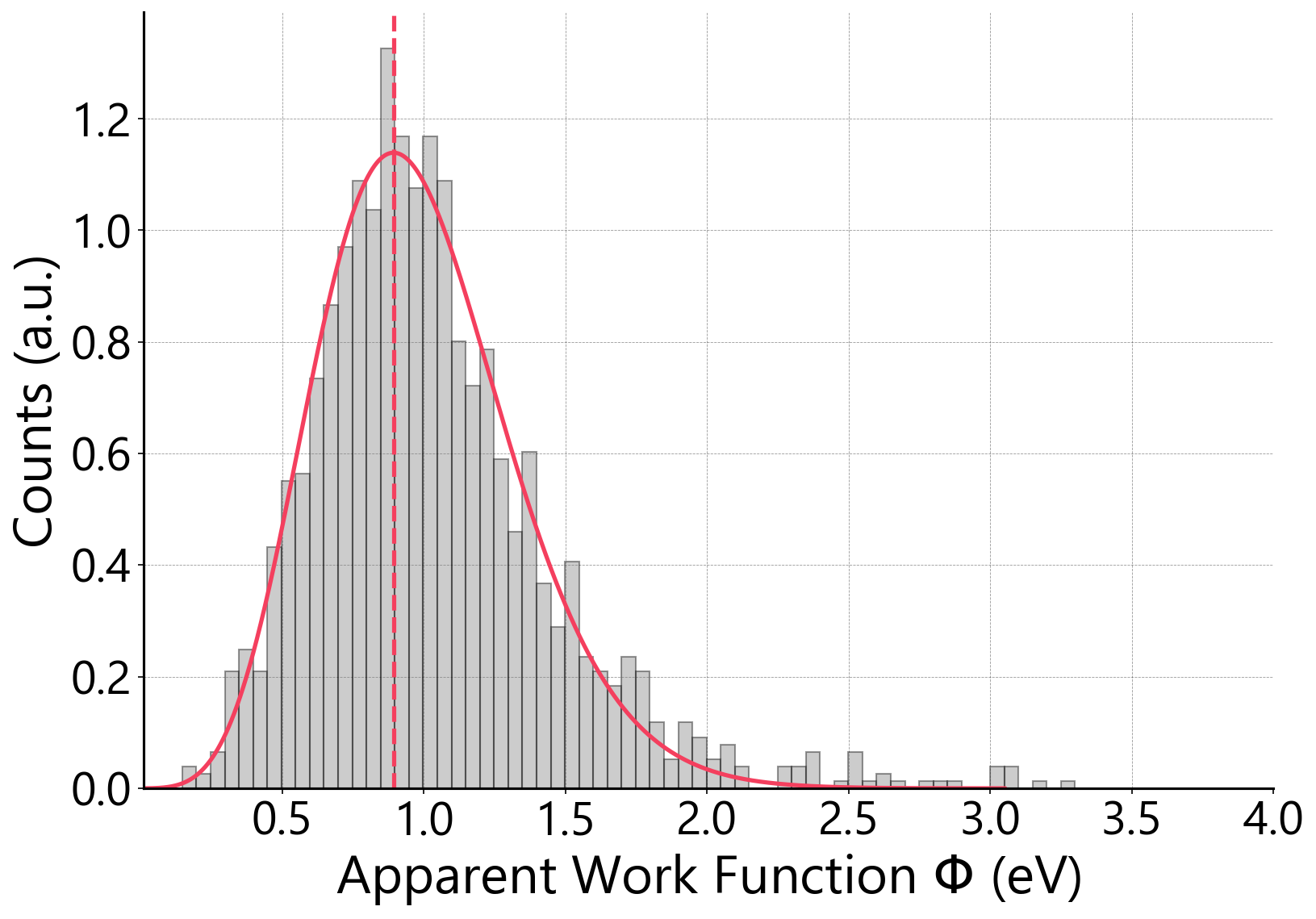}
        
        \put(47, 34){\fbox{\includegraphics[width=0.45\columnwidth]{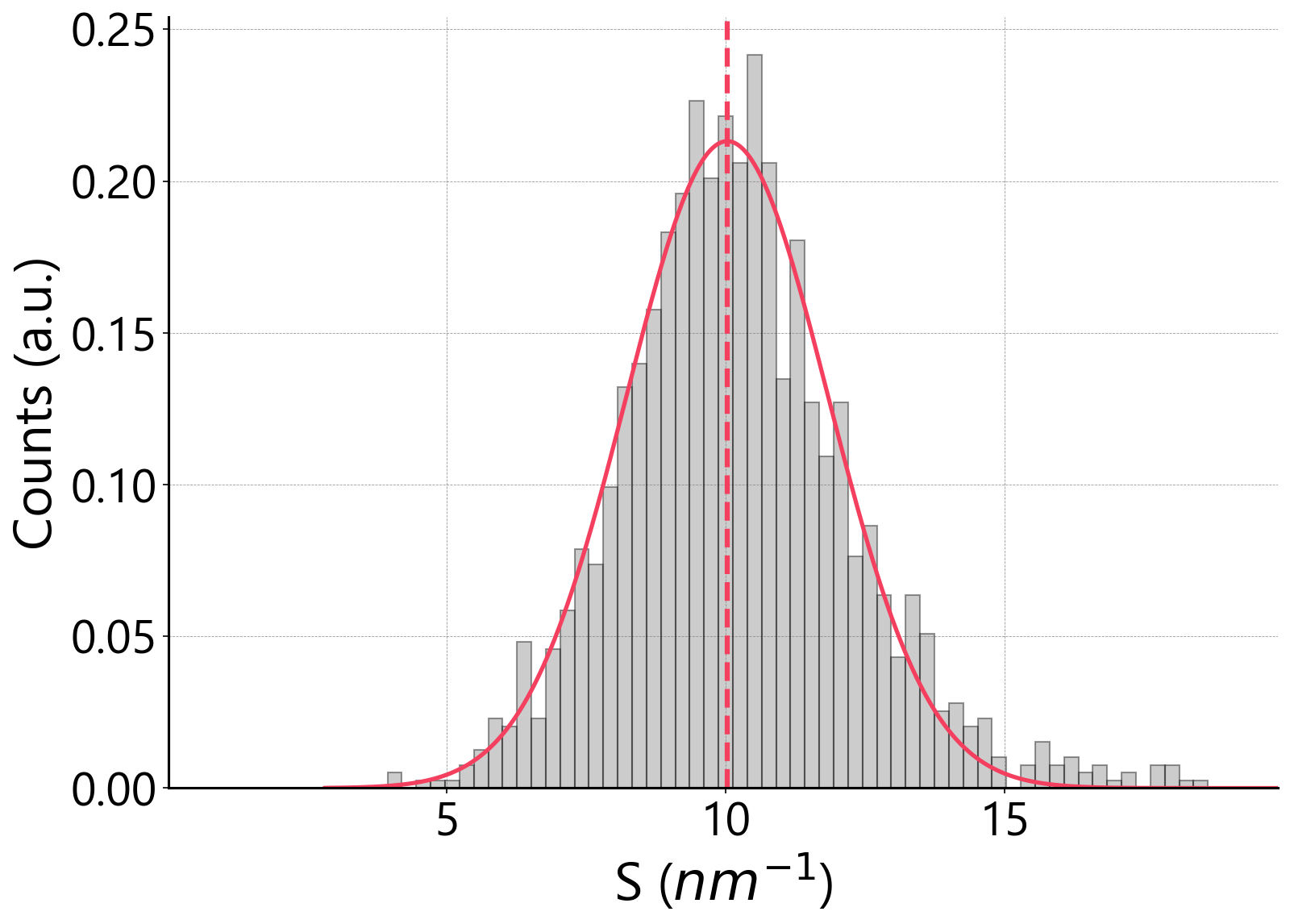}}}
        
    \end{overpic}
    
    \vspace{0.2cm} 
    \caption{\justifying Statistical analysis of the tunneling parameters. The main panel displays the apparent work function histogram fitted with a non-central $\chi^2$ distribution, following our theoretical derivation. To justify this, the inset shows the histogram of the raw tunneling slopes, which, as can be seen, is perfectly fitted by a normal distribution.}
    \label{fig:XiGau}
\end{figure}

To experimentally validate our statistical framework, it is necessary to analyze the probability distributions of the extracted tunneling parameters. As previously discussed, the raw tunneling slopes ($S$) are subject to random experimental atomic configuration, leading them to follow a normal distribution. Because the apparent work function depends quadratically on this slope ($\phi \propto S^2$), fundamental probability theory dictates that its statistical behavior must inherently transform into a non-central $\chi^2$ distribution. Figure \ref{fig:XiGau} directly illustrates this mathematical relationship for a representative dataset: the inset confirms the Gaussian nature of the experimental slopes, while the main panel demonstrates the excellent agreement of the corresponding work function histogram with the predicted $\chi^2$ model.

\subsection{Sensitivity Analysis and Piezoelectric Calibration}

\begin{figure*}[htpb]
    \centering
    \captionsetup[subfigure]{position=top, justification=raggedright, singlelinecheck=false, labelfont=normalsize}

    \begin{subfigure}{0.32\textwidth}
        \caption{} \label{fig:gold_traces}
        \includegraphics[width=\linewidth]{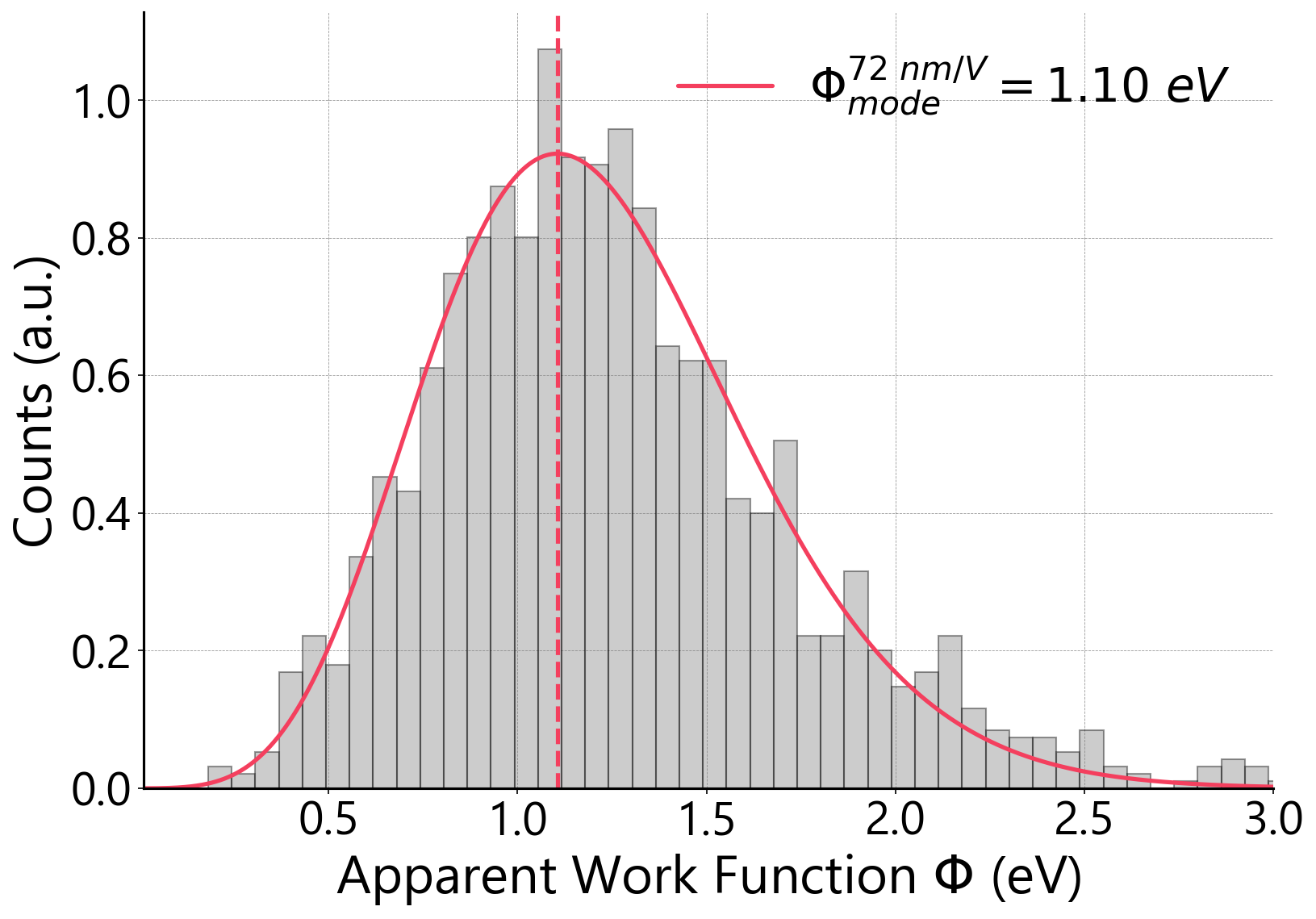}
    \end{subfigure}\hfill
    \begin{subfigure}{0.32\textwidth}
        \caption{} \label{fig:gold_hist}
        \includegraphics[width=\linewidth]{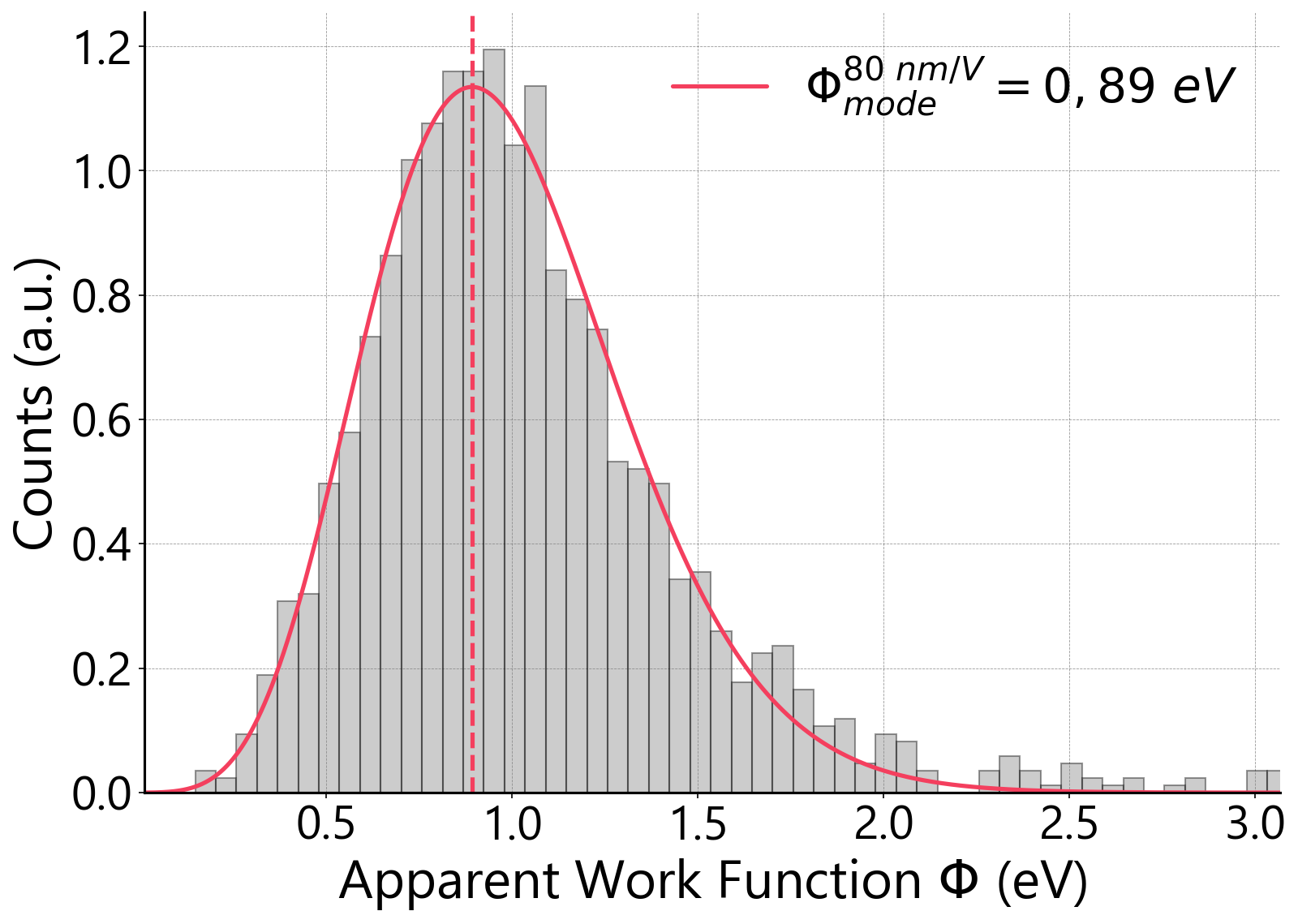}
    \end{subfigure}\hfill
    \begin{subfigure}{0.32\textwidth}
        \caption{} \label{fig:copper_hist}
        \includegraphics[width=\linewidth]{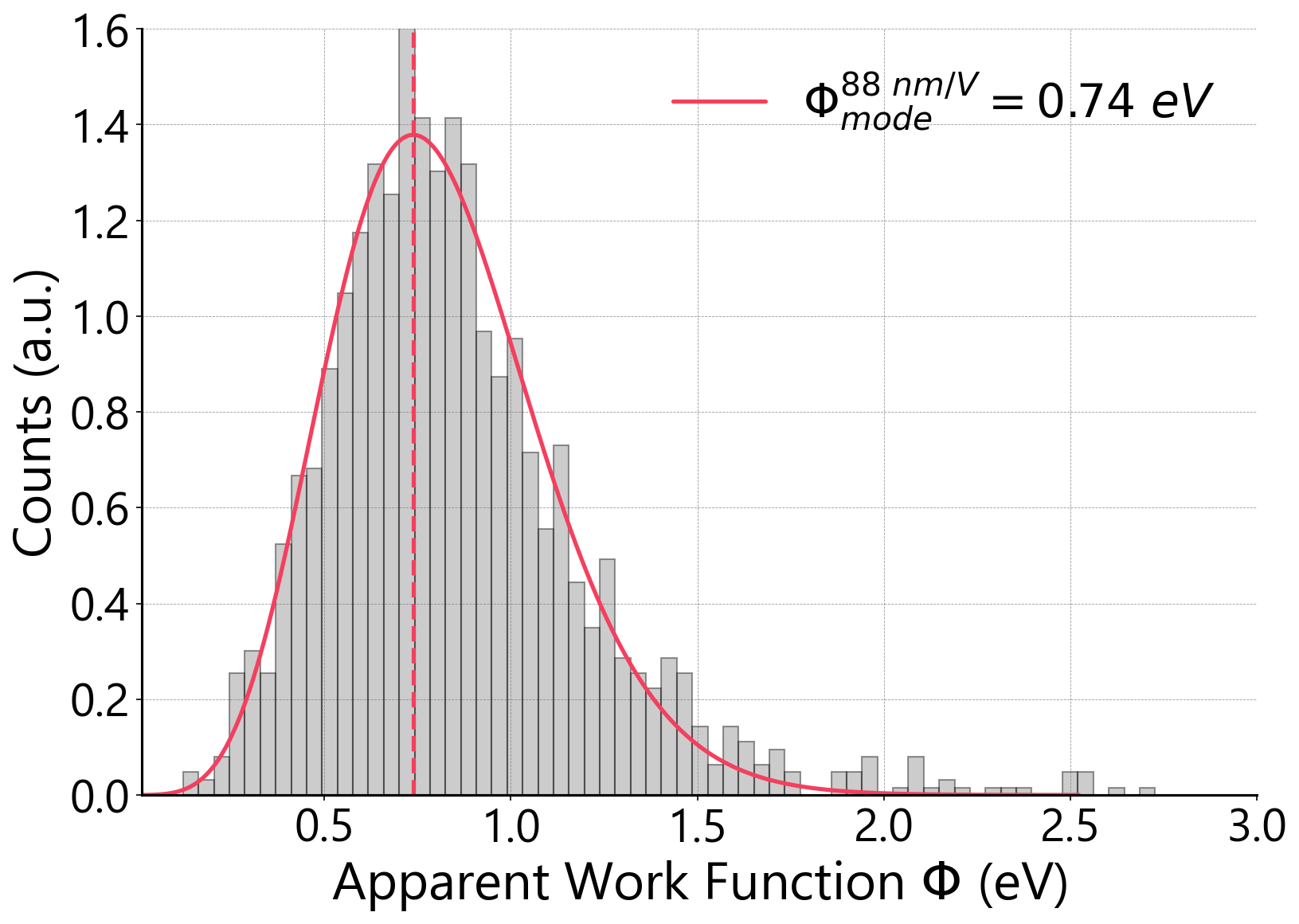}
    \end{subfigure}
    
    \captionsetup[subfigure]{position=top, justification=justified, singlelinecheck=false, labelfont=small}
    \captionsetup{justification=justified, singlelinecheck=false}
    
    \caption{\justifying Impact of piezoelectric calibration uncertainty on the apparent work function. Panel (b) displays the data processed with the nominal calibration factor. For comparison, panels (a) and (c) show the same dataset assuming a $-10\%$ and $+10\%$ calibration error, respectively. This $10\%$ margin corresponds to the typical uncertainty expected when calibrating experimental data using the first three atomic contacts, as reported in \cite{JPCuenca26}.}
    \label{fig:WFSM}
\end{figure*}

Given the quadratic dependence of $\phi$ on the tunneling slope $S$, any uncertainty in the piezoelectric displacement calibration ($z$) propagates quadratically into the work function calculation ($\delta \phi / \phi \approx 2 \cdot \delta z / z$). To quantify this effect, we processed the same raw Au dataset (under room conditions) evaluating three scenarios: a nominal calibration, and deviations assuming a $\pm 10\%$ calibration error. This margin of error is standard in the literature when calibrating displacement using the first three atomic contacts \cite{JPCuenca26}. As shown in Fig. \ref{fig:WFSM} and summarized in Table \ref{tab:calibration_error}, while the absolute position of the distribution peak shifts due to the induced calibration error, the relative shape of the distribution and the physical conclusions regarding environmental degradation remain completely unaffected.

\begin{table}[htpb]
\centering
\caption{\justifying Summary of the $\phi_{\text{mode}}$ for the same Au dataset at room conditions, processed under different piezoelectric calibration factors.}
\label{tab:calibration_error}
\begin{tabular}{@{}lc@{}}
\toprule
\textbf{Calibration Scenario} & \textbf{$\phi_{\text{mode}}$ (eV)} \\ \midrule
Nominal $-10\%$ error & 1.10 \\ 
Nominal calibration   & 0.89 \\ 
Nominal $+10\%$ error & 0.74 \\ \bottomrule
\end{tabular}
\end{table}

\end{document}